\documentclass[3p,final,twocolumn]{elsarticle}

\usepackage{hyperref}

\usepackage{caption}
\usepackage{subcaption}
\usepackage{xcolor}
\usepackage{soul}
\usepackage{amsmath}
\usepackage{float}

\journal{Nuclear Instruments and Methods in Physics Research}                                                                                                                                               

\journal{}
\begin{document}
\begin{frontmatter}

\title{Laser Calibration System for \\ Time of Flight Scintillator Arrays}                                                                                                                                                                       

\author{A.~Denniston}
\author{E.P.~Segarra}
\author{A.~Schmidt\corref{mycorrespondingauthor}\fnref{gwu}}
\ead{schmidta@mit.edu}
\author{A.~Beck\fnref{nrcn}}
\author{S.~May-Tal Beck\fnref{nrcn}}
\author{R.~Cruz-Torres}
\author{F.~Hauenstein\fnref{odu}}
\author{A.~Hrnjic}
\author{T.~Kutz\fnref{gwu2}}
\author{A.~Nambrath}
\author{J.R.~Pybus }
\author{P.~Toledo}
\author{O.~Hen}
\address{\MIT}

\author{L.B.~Weinstein}
\address{\ODU}

\author{M.~Olivenboim}
\author{E.~Piasetzky}
\address{\TAU}

\author{I.~Korover}
\address{\NRCN}

\cortext[mycorrespondingauthor]{Corresponding Author}
\fntext[gwu]{Present address: \GWU}
\fntext[nrcn]{Also at: \NRCN}
\fntext[odu]{Also at: \ODU}
\fntext[gwu2]{Also at: \GWU}

\newcommand*{\MIT }{Massachusetts Institute of Technology, Cambridge, Massachusetts 02139, USA}
\newcommand*{\GWU }{George Washington University, Washington, D.C.  20052, USA}
\newcommand*{\ODU}{Old Dominion University, Norfolk, Virginia  23529, USA}
\newcommand*{\TAU}{School of Physics and Astronomy, Tel Aviv University, Tel Aviv  69978, Israel}
\newcommand*{\NRCN}{Nuclear Research Center Negev, Beer-Sheva  84190, Israel}

\begin{abstract}

A laser calibration system was developed for monitoring and calibrating time of flight (TOF) scintillating detector arrays.  The system includes setups for both small- and large-scale scintillator arrays.  Following test-bench characterization, the laser system was recently commissioned in experimental Hall B at the Thomas Jefferson National Accelerator Facility for use on the new Backward Angle Neutron Detector (BAND) scintillator array.  The system successfully provided time walk corrections, absolute time calibration, and TOF drift correction for the scintillators in BAND.  This showcases the general applicability of the system for use on high-precision TOF detectors.

\end{abstract}


\end{frontmatter}

\section{Introduction}
    
Time of flight (TOF) detectors are used to measure the time it takes for a particle to travel a given path length.  This can provide particle identification by measuring the TOF of particles with a known momentum, determine the momenta of known particles, or determine the coincidence of multiple particles detected from a single interaction.  Typically, TOF detectors consist of a plane of scintillating detectors whose signal arrival time can be compared to a reference time.

Large-scale TOF detectors often use many separate scintillators to cover large solid 
angles with good spatial granularity.  Each scintillator requires dedicated light
detectors and readout electronics.  This in turn requires the precise absolute and
relative calibration of all scintillators in the detector. While data from 
over-constrained reaction channels can be used for these calibrations, a 
dedicated calibration system based on an external light source is often
more flexible, more convenient, and more precise. A pulsed laser system that
delivers simultaneous light pulses to each scintillator with known amplitude 
can quickly establish a full set of calibrations and correct any time-dependent
drifts. 

\begin{figure*}[t]
\centering
\includegraphics[width=0.75\textwidth]{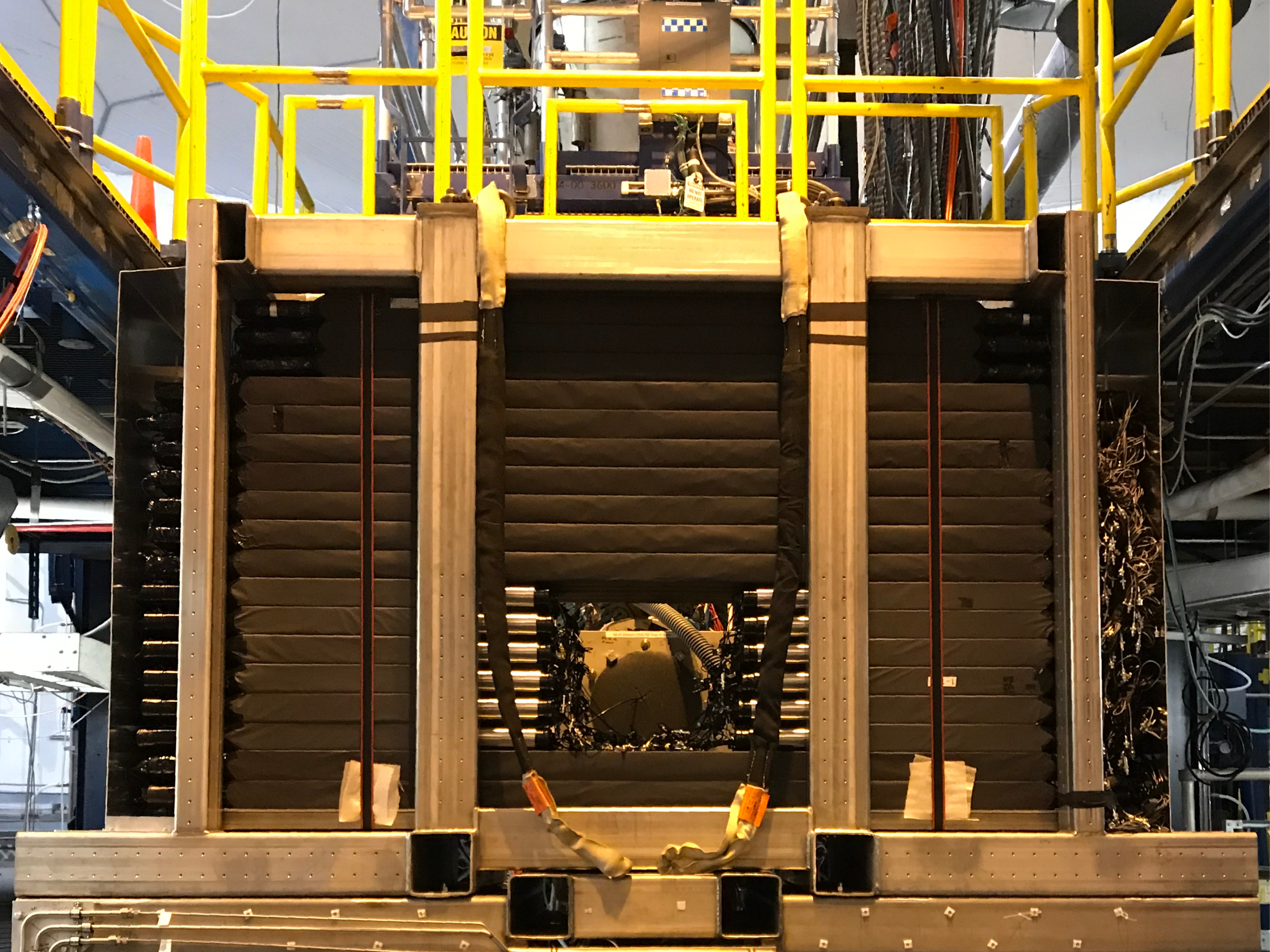}
\caption{The backward angle neutron detector (BAND) installed in Hall B at Jefferson Lab.}
\label{fig:bandpic}
\end{figure*}

Such a laser system was developed for the backward angle neutron detector (BAND),
a TOF array consisting of 140 scintillator bars \cite{band}.   BAND, pictured in
Figure~\ref{fig:bandpic}, was recently deployed in experimental Hall B at the
Thomas Jefferson National Accelerator Facility (JLab).  BAND was designed to detect
backwards recoiling neutrons over the momentum range of 200--600~MeV$/c$, specifically
spectator neutrons from the deep inelastic of electrons off of protons bound in 
deuterium. This technique of spectator-tagged DIS allows the determination of the
proton's nuclear modification as a function of virtuality, with the goal of
elucidating the relationship between the EMC Effect and the short-range correlations
between nucleons~\cite{Hen:2016kwk}. BAND's design goal is to measure neutron time of flight with a 
resolution better than 300~ps, making timing calibrations at the level of 100~ps paramount.

The required specifications for the calibration of TOF arrays, including but not limited to BAND,
are discussed in the following section. The laser system discussed in this article provided
critical calibrations for BAND and served as a proof of concept for application in other large
TOF scintillator arrays.

\section{Design criteria}

The exact requirements for the calibration of scintillator arrays depend on the specific demands of the detector.  However, some general assumptions are valid for most plastic scintillating detectors.  Their time resolution should be at the level of a few hundreds picoseconds (ps) or better. Therefore, the calibration system should provide short pulses with a pulse width up to a few nanoseconds (preferably $<$1 ns) and a rise time on the order of tens of picoseconds.  Similar considerations apply also to the amplitude calibration, where a resolution of 5-10\% is required for the time walk correction and a systematic study of discriminator thresholds.

To best emulate the excitation of scintillators by charged particles, we use ultraviolet (UV) radiation of 355 nanometers (nm) where individual photons are energetic enough to excite the scintillant molecules.  As the UV region is not widely used by the optical engineering community, special considerations were necessary to accommodate long-term use with minimal radiation damage and continuous degradation of the optical components.  The optical components were chosen to be easily replaceable and low-cost (few hundreds of dollars). 

The use of UV light leads to safety issues. These can be minimized by using a fiber-coupled solid-state laser and a high-efficiency light transport system. The latter minimizes the required laser power while still delivering sufficient light to individual scintillators.
Reducing the laser power also minimizes the possible degradation of optical components.

A fiber distribution system is required to deliver laser pulses with common timing and similar amplitude to each scintillator bar with minimal light loss. While such laser distribution systems (i.e. splitters) are common, the assumed maximum detector array size of 400 scintillators required a custom system.  

To facilitate moving the system between multiple experiments, it was designed compact, portable, and modular.  This allows the optical components to be easily customized for different experimental demands.  This is made possible by the small size of solid-state lasers and individual components.  

Lastly, it was crucial that operation of the laser system be inherently safe to comply with safety regulations of nuclear research facilities.
    
\begin{figure*}
\begin{subfigure}{\textwidth}
\centering
\includegraphics[width=0.85\textwidth]{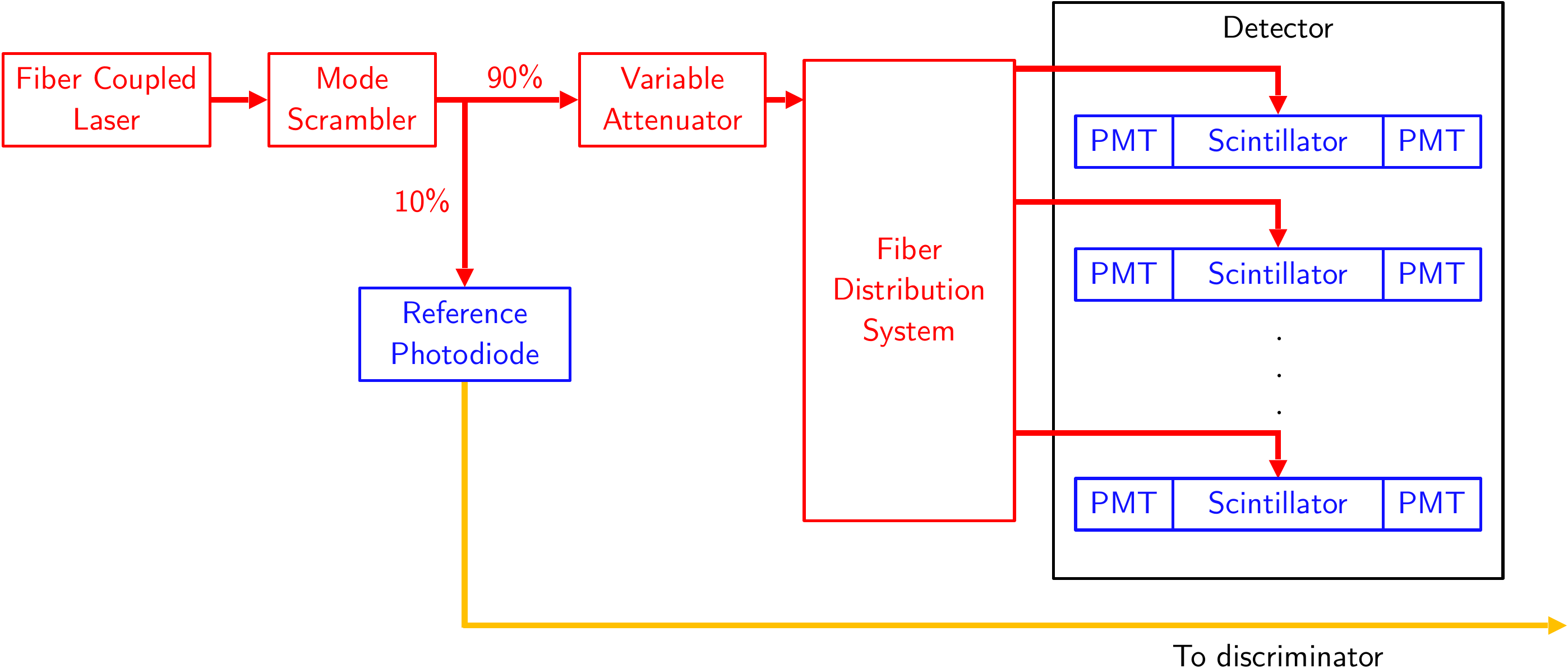}
\caption{Laser system}
\label{fig:oSetup}
\end{subfigure}
\vspace{25pt}
\begin{subfigure}{\textwidth}
\centering
\includegraphics[width=0.85\textwidth]{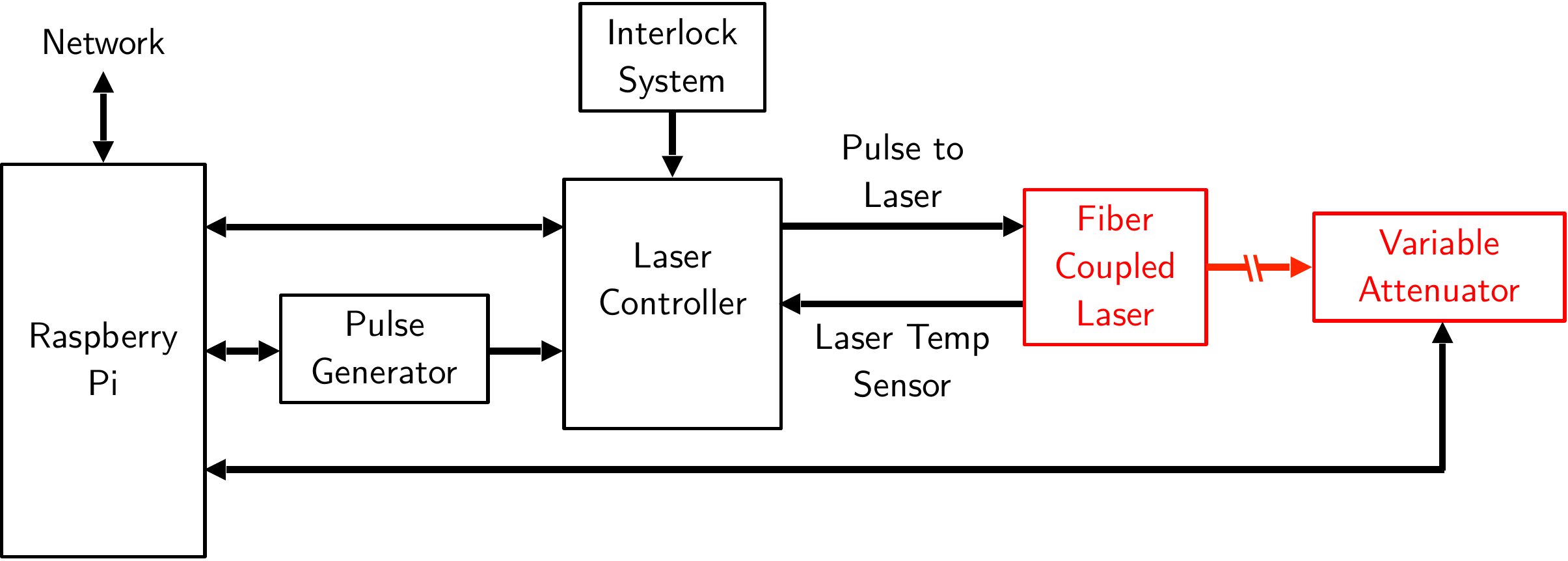}
\caption{Laser controller system}
\label{fig:cSetup}
\end{subfigure}

\vspace{25pt}

\begin{subfigure}{\textwidth}
\centering
\includegraphics[width=0.85\textwidth]{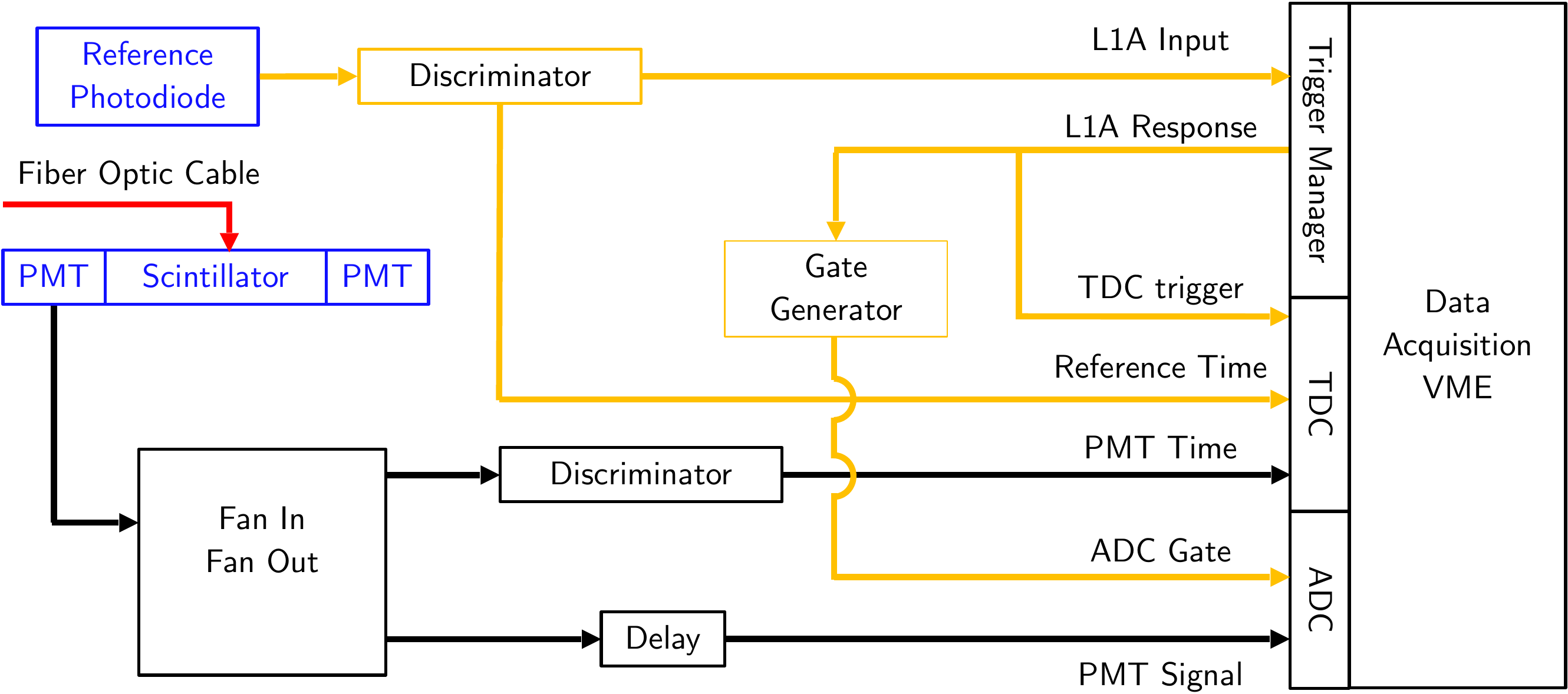}
\caption{Electronics setup}
\label{fig:eSetup}
\end{subfigure}

\caption{Schematic diagrams of the laser system, laser controller, and electronics.  Red indicates optical components.  Black indicates electronic components.  Orange indicates electronic components downstream from the reference photodiode.  Blue indicates detectors.}

\end{figure*}

\begin{figure*}[t]
\centering
\includegraphics[height=0.5\textwidth]{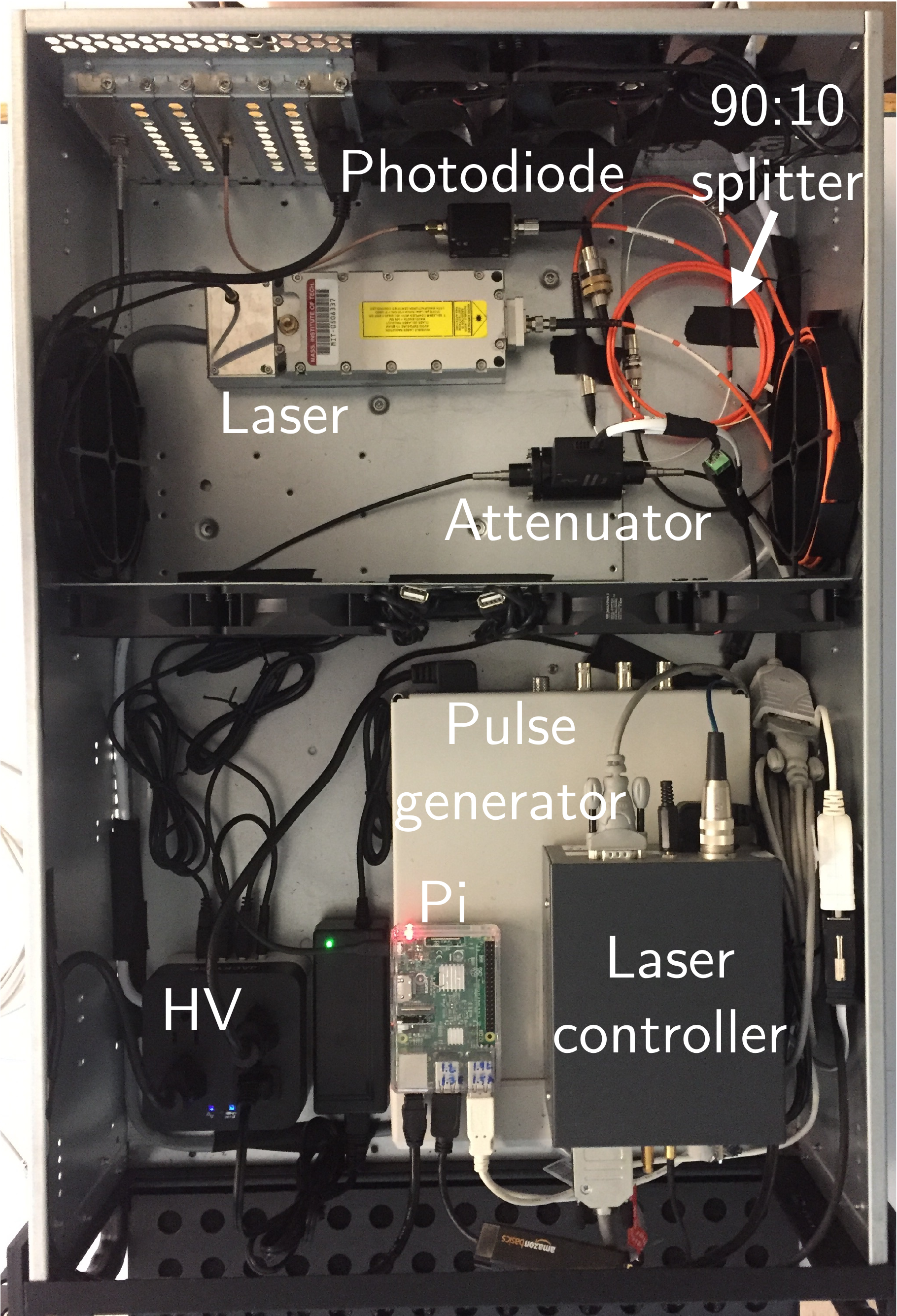}
\hspace{10pt}
\includegraphics[height=0.5\textwidth]{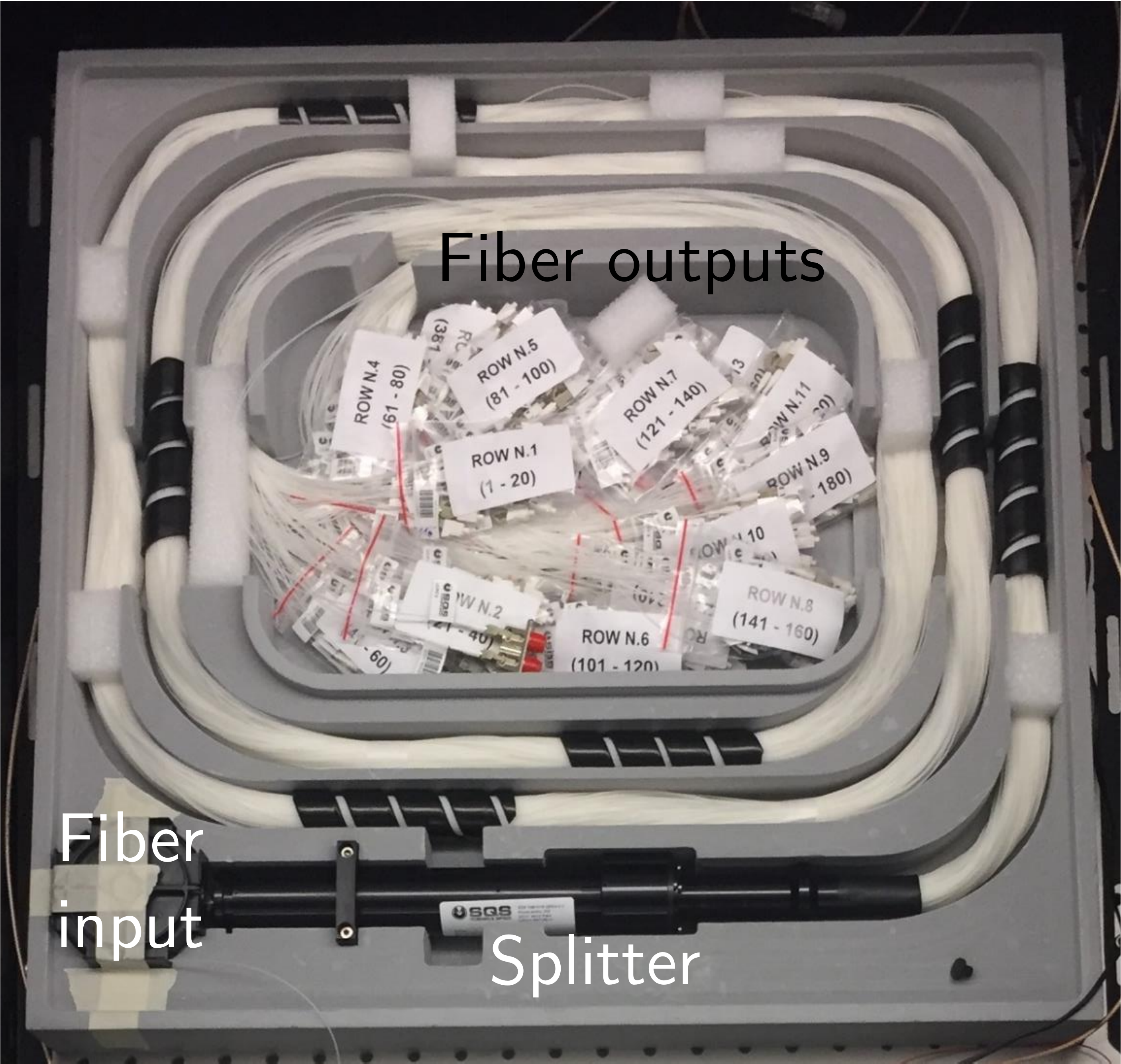}
\caption{The laser system in its rack-mountable case (left).  The 1$\times$400 fiber splitter (right).}
\label{fig:system}
\end{figure*}

\section{Laser System}
    
\subsection{Source laser} 

The primary laser chosen for the system is a Teem Photonics STV-01E-140 Nd:YAG pulsed laser, producing light with a wavelength of 355 nm in a single longitudinal mode \cite{teem_laser}.  The laser is optimized to create pulses less than 400 ps in duration, with a repetition rate between 10 Hz and 4 kHz.    

The laser couples directly to a fiber-optic cable, and the pulses are contained within fiber-optic cables throughout the entire system until entering the scintillators.  This laser complies with the three aforementioned demands: (1) UV wavelength to mimic scintillation from ionizing radiation, (2) short rise-time of the laser pulse, and (3) closed optical route for safety.

Each laser pulse has an energy of 1 microjoule ($\mu$J), significantly smaller than that required by
free-space gas lasers, like those employed in the TOF calibration systems of the former
CLAS \cite{Smith:1999ii} and BLAST \cite{Hasell:2009zza} experiments. Even after splitting the light
into 400 output fibers, this was still more than enough per pulse to mimic the light response of 
scintillators to high-energy particles. Test bench studies showed that a laser pulse of 25~pJ
was approximately equivalent to the light produced by cosmic rays in a BAND scintillator module. 

A second, less expensive, 405~nm Thorlabs NPL41B pulsed laser \cite{thor_laser} was also used for
initial testing and comparative table-top studies.  The laser was tested to ensure the wavelength
was sufficiently short to excite standard plastic scintillators.  The laser pulses have a minimum
width of 6~ns, with a repetition rate of up to 10 megahertz (MHz).  Each pulse has an energy of
1.5~nJ, several hundred times smaller than the primary 355 nm laser.  This laser is not fiber-coupled,
requiring the implementation of a Thorlabs PAF-X-18-PC-A FiberPort coupler \cite{thor_coupler}. 
Two Thorlabs NB1-K08 mirrors \cite{thor_mirror} are used to ensure that the incoming laser is
aligned with the fiber axis of the coupler. 

Both lasers are single-mode, which is incompatible with the fiber distribution system described below.
We therefore added a Newport FM-1 fiber mode scrambler at the laser exit to produce a stable multi-mode
distribution from the single-mode laser pulse \cite{newport_mode}. 

\subsection{Laser controller}

The primary laser is operated using a Teem Photonics MLC-03A-DP1 laser controller \cite{teem_control} and a Siglent SDG 1032X pulse generator \cite{siglent_pulse}, which serves as the trigger for the laser controller.  

Both the laser controller and pulse generator are controlled by a Raspberry Pi computer that also monitors the temperature readout from the laser itself.  The laser controller's safety interlock system prevents it from being energized if the laser enclosure is open.  The setup is depicted in Fig. \ref{fig:cSetup} and~\ref{fig:system}.

\begin{figure*}[h]
\centering
\includegraphics[width=1.0\textwidth]{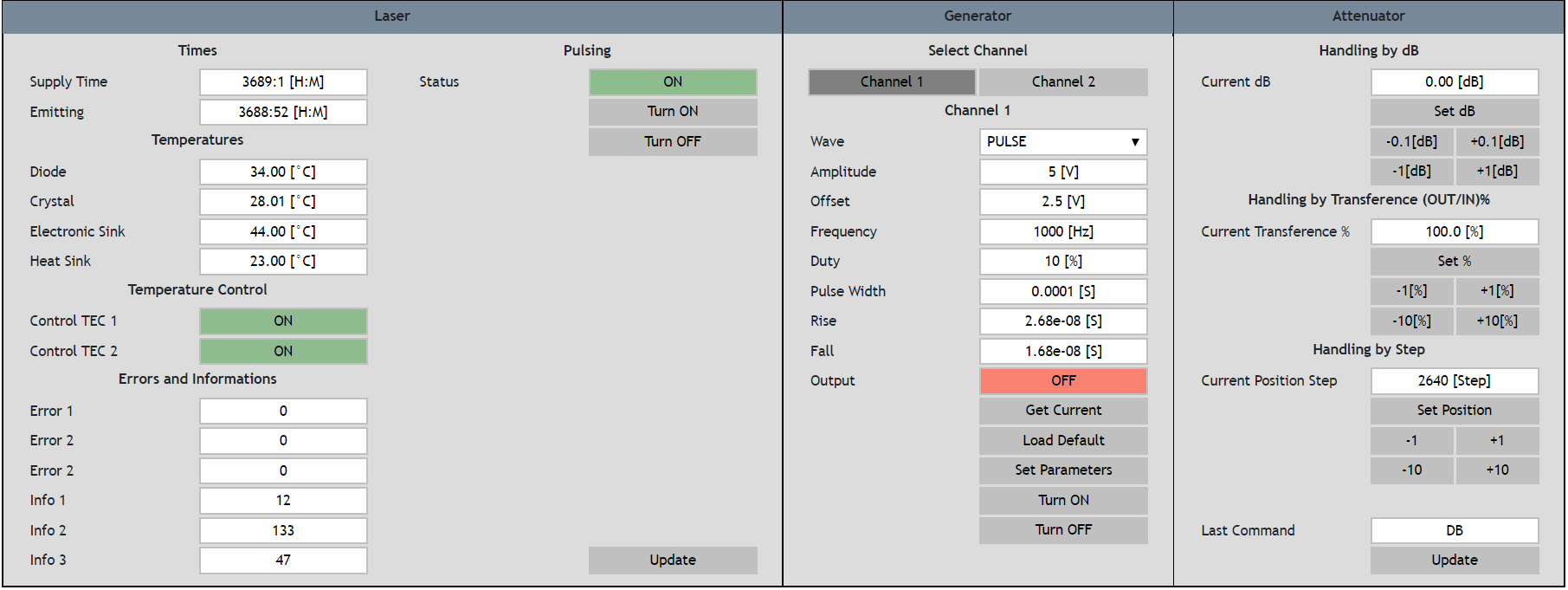}
\caption{GUI for remote control of the laser system. These panels show settings for the laser controller,
the pulse generator, and optical attenuator, which can be adjusted remotely, without any need to enter
the experimental area.}
\label{fig:webGUI}
\end{figure*}

To control the laser system, the Pi connects to the network via Ethernet and provides a GUI through an HTTP server (see Figure~\ref{fig:webGUI}).  This allows the Pi to directly connect to the Internet for remote control. The GUI allows control of the trigger pulse settings, the attenuator level, monitoring the laser temperature, and powering up/down the entire system.

\subsection{Optical setup} 
After a light pulse leaves the laser in an optical fiber, it is processed using various optical components that split and distribute the light, as shown in Figure~\ref{fig:oSetup} and discussed below. 

An initial 90:10 fiber splitter is used to divide the laser pulse into a reference pulse and a pulse to be distributed to the detector.  

The 10\% reference pulse is sent to a Thorlabs DET025AFC/M fiber-coupled silicon photodetector \cite{thor_photo}.  The signal from this photodetector is used for the TDC reference time and trigger (see Section~\ref{sec:detector}).  

The remaining 90\% of the pulse is sent to a custom motorized attenuator by Oz Optics~\cite{oz_attenuator}. The attenuator is used to vary the light pulse intensity sent to the detectors. It is operated by a dedicated 12V power supply and consists of two blades with a precise ultrasonic computer-controlled actuator, providing nearly continuous transmission from 100\% to 1\%.  The attenuated pulse is then sent to the fiber distribution system.

\subsection{Fiber distribution system}
\label{sec:lds}

To distribute the laser pulse to each scintillator bar with similar amplitude and common timing, a custom SQS Vl\'aknov\'a Optika 1$\times$400 splitter is used.  The splitter contains a series of three multilens arrays (MLAs).  The first two shape the incoming laser pulse to a tophat profile.  The third MLA focuses this tophat beam into each of the 400 output fibers, which are equipped with LC connectors.  The splitter is shown in Figure~\ref{fig:system} (right).

\begin{figure*}[h]
\begin{subfigure}{0.5\textwidth}
\centering
\includegraphics[width=\columnwidth]{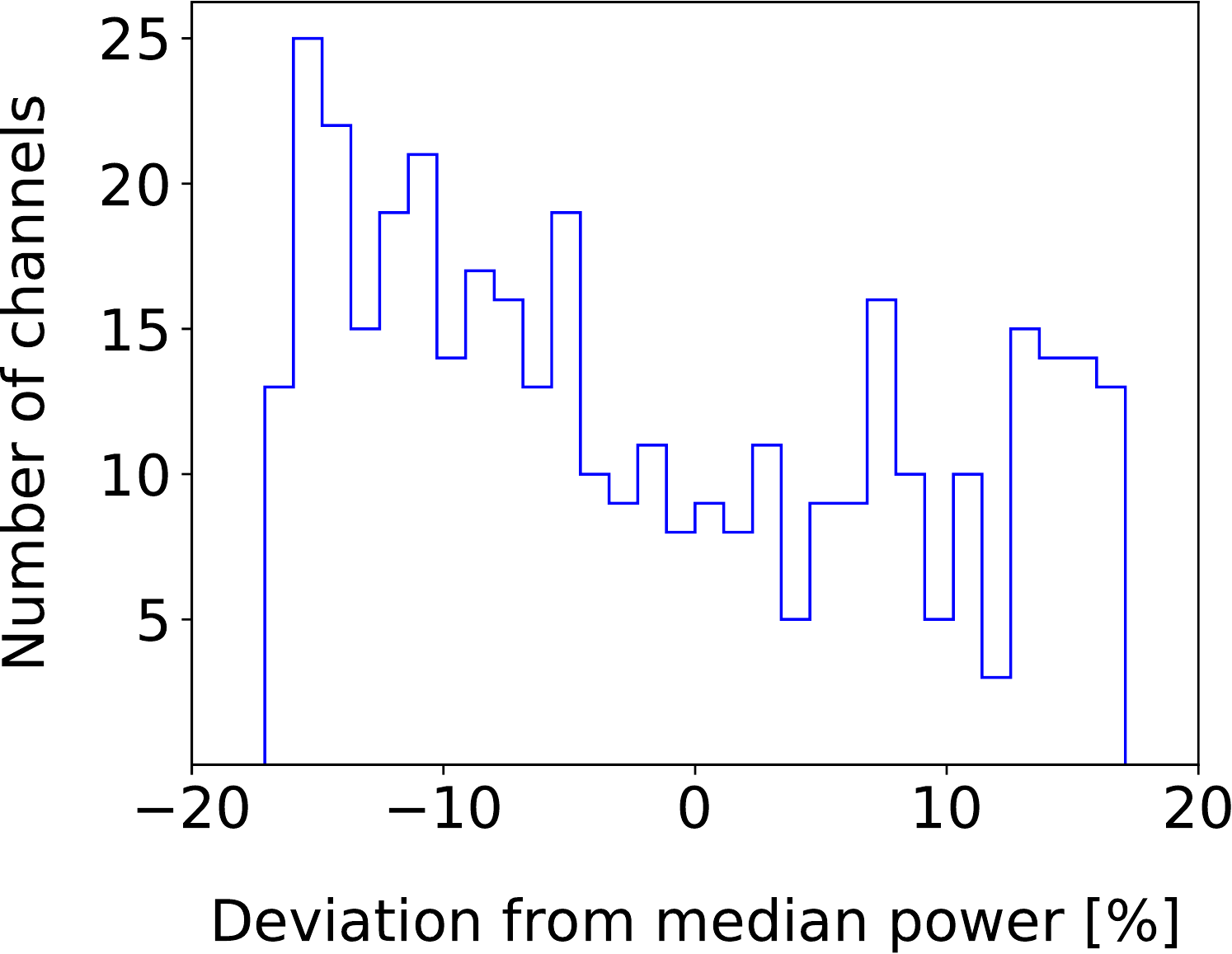}
\caption{}
\label{fig:splitterAmp}
\end{subfigure}
\begin{subfigure}{0.5\textwidth}
\centering
\includegraphics[width=\columnwidth]{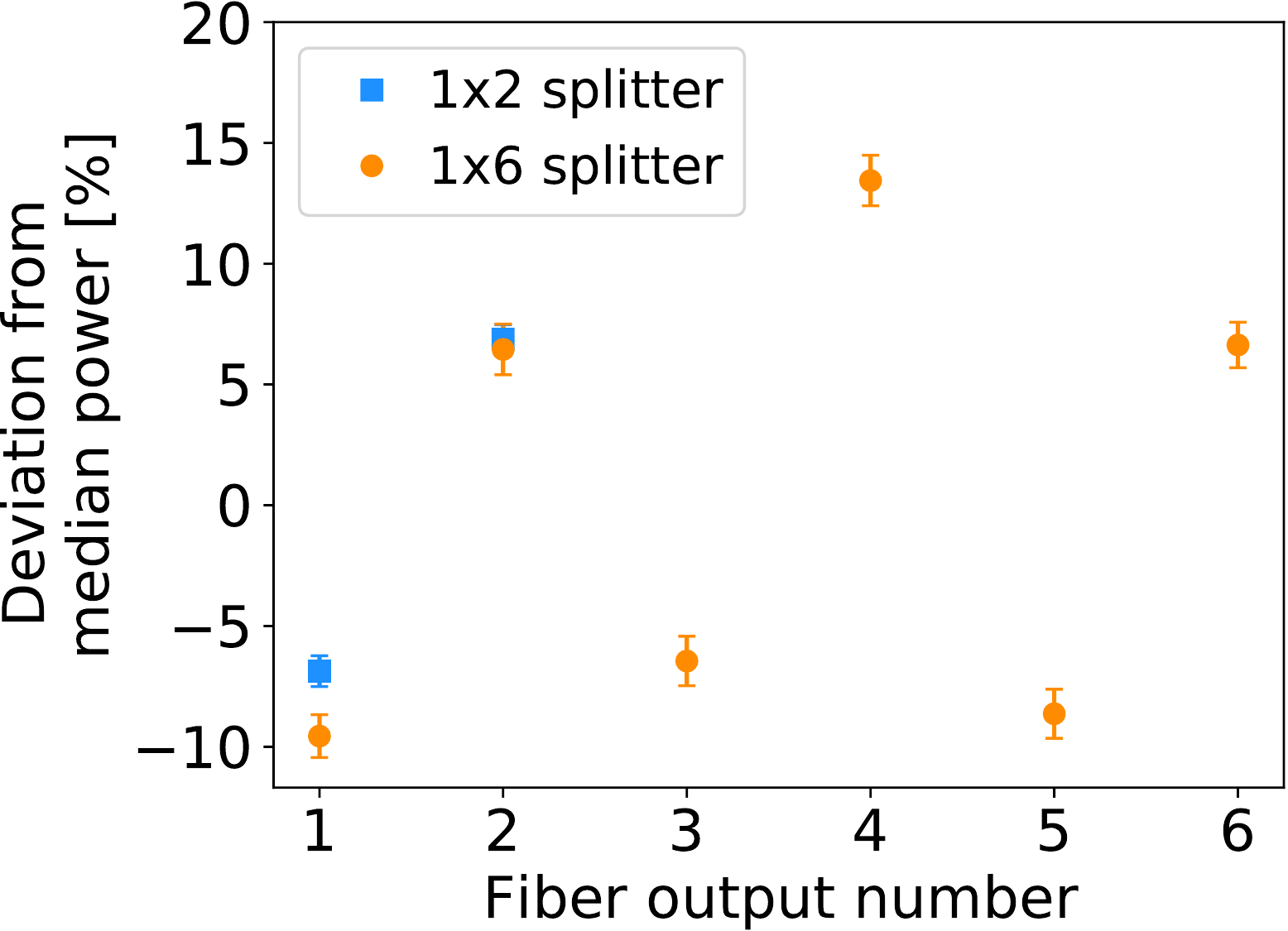}
\caption{}
\label{fig:1x6}
\end{subfigure}
\caption{(a) Power distribution of the 1$\times$400 splitter output fibers.  The horizontal axis indicates percent deviation from the median power.  (b) Power distribution of the small-scale splitter output fibers.  The vertical axis indicates percent deviation from the median power.}
\end{figure*}

The uniformity of the splitter was tested by the manufacturer using a 405 nm laser.  The resulting power distribution is shown in Figure~\ref{fig:splitterAmp} as the percent deviation from the median power.  Over 95\% of the fibers had less than 20\% variation in power from the median fiber, meeting the stability guaranteed by the manufacturer.  Fibers outside of this range were not used for the calibration system and are not included in Figure~\ref{fig:splitterAmp}.

An Lfiber 1$\times$6 splitter \cite{lfiber_splitter} was also tested for use on smaller-scale detectors.  For reference, a Thorlabs 1$\times$2 (50:50) splitter \cite{thorlabs_splitter} was tested with the same setup.  

The uniformity of the 1$\times$6 splitter was estimated using the primary laser.  The test used a minimal set-up, which included only the laser, attenuator, splitter, and photodiode.  A reference measurement without the splitter resulted in the laser instability of about 2\%. Figure~\ref{fig:1x6} shows the percent deviation from median power for each fiber, as measured for the 1$\times$6 and 1$\times$2 splitters.  Each data point is the average of repeated measurements.  The 1$\times$2 splitter shows a deviation of $\pm$7\%, consistent with the manufacturers tolerance of 5\% (when the laser instability is considered).  The 1$\times$6 splitter shows larger deviations between -10\% and 15\%. 

\subsection{Coupling to scintillators}

After the splitter, the light from each fiber needs to be directed into the individual detectors.
Geometrical restrictions usually prevent placing the fiber such that if shines directly into the face of the scintillator.
Specifically for BAND, following previous work~\cite{Smith:1999ii}, the fiber is placed parallel to the scintillator between its surface and the optical reflector wrapped around the bar ~\cite{band}. 

Several different fiber terminations were tested in an attempt to optimize the light yield and stability of pulses entering the detector.  These included:
\begin{itemize}
    \item Fiber severed at 90$^\circ$ relative to cable direction.
    \item Fiber severed at 45$^\circ$  relative to cable direction (so-called ``side fire" termination)
    \item Cylindrical diffuser tip used to diffuse light uniformly over several centimeters. 
    \item Spherical diffuser tip used to diffuse light from a localized point.
    \item Fiber exposed (no jacket or cladding) for last few centimeters prior to termination, allowing light to escape the fiber.
\end{itemize}
The various termination methods were tested using the secondary laser and attenuator.  The attenuated laser pulse was split using the 1$\times$2 splitter discussed in Section~\ref{sec:lds}.  One output was directed to a reference PMT mounted inside a dark box.  The other was connected to the fiber being tested, which was attached to the scintillator.  Multiple orientations of the scintillator fiber were tested, such as reversing the direction of the fiber tip and rotating the fiber about its axis.  The attenuator transmission was varied from 0\% to 4.5\%, the maximum range allowed without saturating the PMTs.  

\begin{figure*}[h]
\begin{subfigure}{0.5\textwidth}
\centering
\includegraphics[width=\columnwidth]{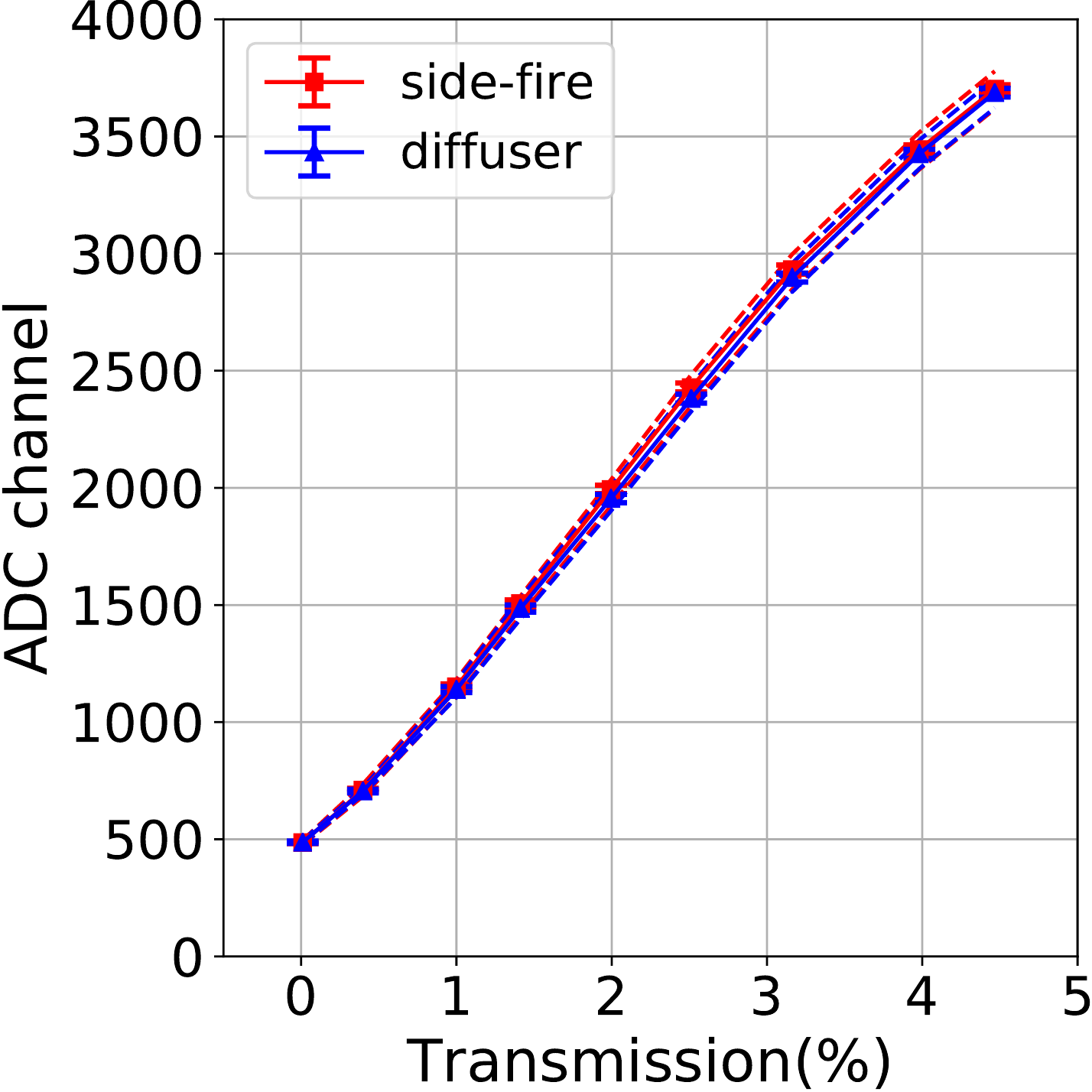}
\end{subfigure}
\begin{subfigure}{0.5\textwidth}
\centering
\includegraphics[width=\columnwidth]{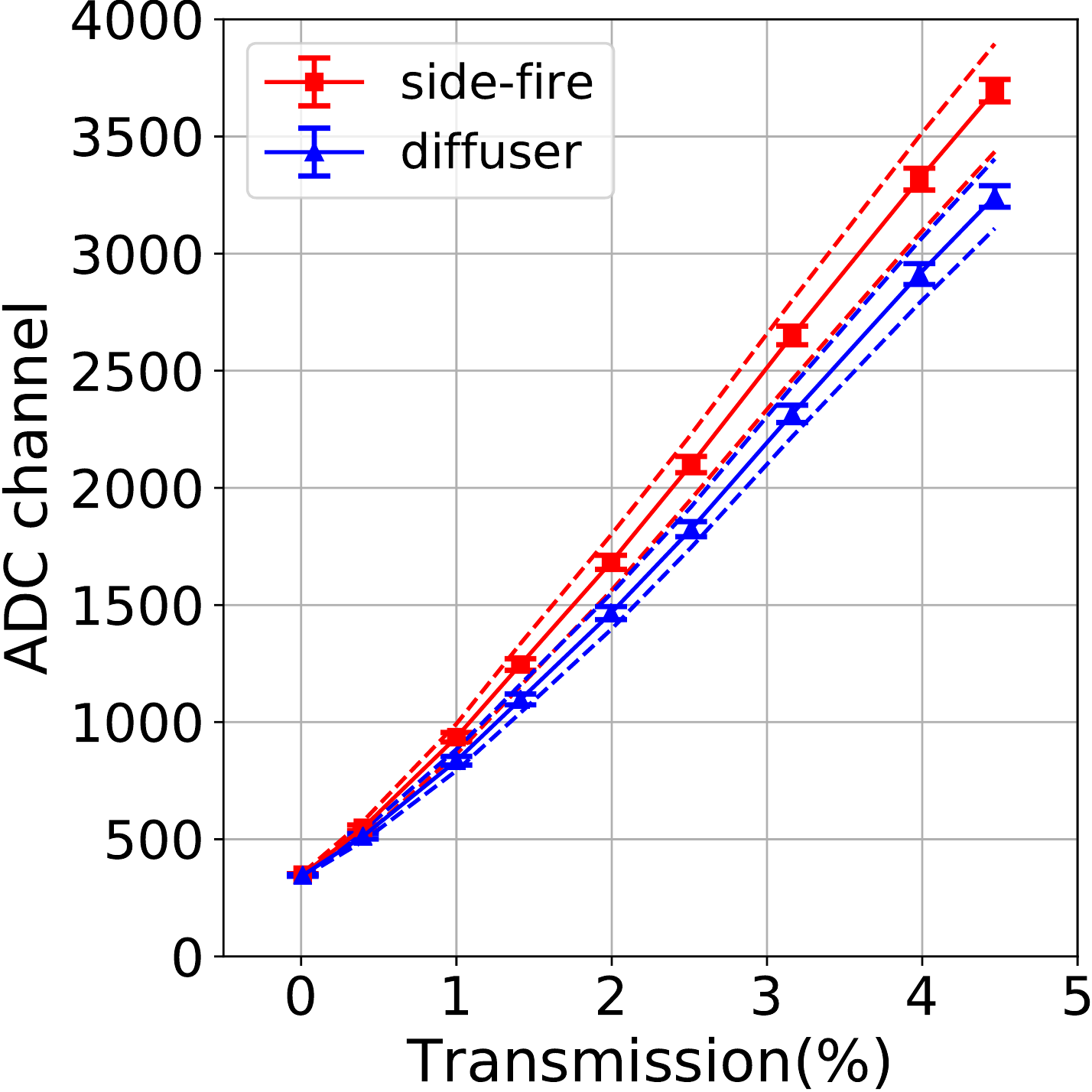}
\end{subfigure}
\caption{Comparison of PMT measurements for the side fire and cylindrical diffuser terminations for the reference PMT (left) and scintillator-coupled PMT (right).  Points represent the average of multiple measurements, with dashed lines indicating the minimum and maximum measurements used in the average.}
\label{fig:fiberEnd}
\end{figure*}

A comparison of these measurements for the side fire and cylindrical diffuser are shown in Figure~\ref{fig:fiberEnd}, where each data point represents the average of the different fiber orientations.  The dashed lines indicate the minimum and maximum of the averaged measurements for each transmission setting.  As expected, the measurements made with the reference PMT are nearly identical for both termination methods, demonstrating stable performance of the optics system during the measurements.  The asymmetric side fire termination has the largest sensitivity to the fiber orientation, resulting in a larger spread compared to the symmetric cylindrical diffuser.  Both terminations show linear response for transmission in the range of 1\% to 4.5\%.  The remaining termination methods showed similar behavior.   

As the exposed fiber method was the most stable, as well as the easiest and cheapest to implement, it was chosen for BAND.  A cut was made in the optical reflector at the center of the scintillator, and the fiber end was inserted into this cut and fixed to the scintillator with RTV615 silicon glue.  The bar was then wrapped in a light-tight material.

\section{Scintillator test setup}

\label{sec:detector}

A test-bench scintillator detector was set up to test and optimize the performance of the system. It consisted of a typical BAND scintillator~\cite{band} with dimensions of $7.2 \times 7.2 \times 200$ cm$^3$, with an optical fiber attached as discussed above. It is read out by two Hamamatsu R7724 PMTs \cite{hamamatsu_pmt} positioned on both ends of the scintillator.

The electronics setup used to process the PMT signals is shown in Figure \ref{fig:eSetup}.  It consists of a LeCroy 428F linear fan-in fan-out (FIFO) module, a LeCroy 623B octal discriminator, a Philips Scientific 794 gate generator, and finally a CAEN V1290A TDC and CAEN V792 ADC \cite{lecroy_fifo, lecroy_disc, phillips_gate, caen_tdc, caen_adc}.  The TDC and ADC are in a VME crate that interfaces the electronics with a computer recording the data. 

The photodiode reference signal is sent to the FIFO module, which provides one output to the ADC and the other to the discriminator.  The discriminator outputs are used for the TDC reference time and the Level One Accept (L1A) of the VME trigger manager.  The L1A output serves as a common stop for all TDC channels and triggers the gate generator to produce the gate signal for the ADC.  

The signals from the scintillator PMTs are also sent to the FIFO modules, allowing the same PMT signal to be processed for both the TDC and ADC.  One FIFO output is sent to the discriminator, with the output sent to the TDC.  This allows the signal arrival time of each bar to be compared to the arrival time of the reference signal.  During calibration with the laser system, the time difference $\Delta t$ between the signal from the PMT and the signal from the reference photodetector is measured:
\begin{equation}
\label{eq:tof}
\Delta t = t_{PMT} - t_{ref}
\end{equation}

A second FIFO output is sent to the ADC, with a delay added such that the signal coincides with the ADC gate.  The ADC signal allows a determination of the energy deposited by the signal, used for a time walk correction (see discussion below).  The digitized TDC and ADC signals are recorded on the computer for offline analysis.  

\subsection{Filter tests}

As the sensitivity of the PMTs extends below 300 nm, they are capable of detecting light directly 
from the laser itself in addition to the induced scintillation light. This effect would be problematic
for calibrations, since it would lead to a difference between the PMT response to laser-induced and particle-induced
scintillation light. 

We tested this effect using a 400~nm longpass filter, which blocked light from the laser, but transmitted light
produced by scintillation. We compared the PMT signals from a short 5~cm scintillator with and without the filter.
We found no detectable difference, implying that the PMTs were responding to scintillation light, rather than direct
laser light. A similar test was carried out using the 405~nm laser and a 410~nm longpass filter, with the same results.

\section{Performance}

\subsection{Time resolution}

\begin{figure}[b!]
\centering
 \includegraphics[width=\columnwidth]{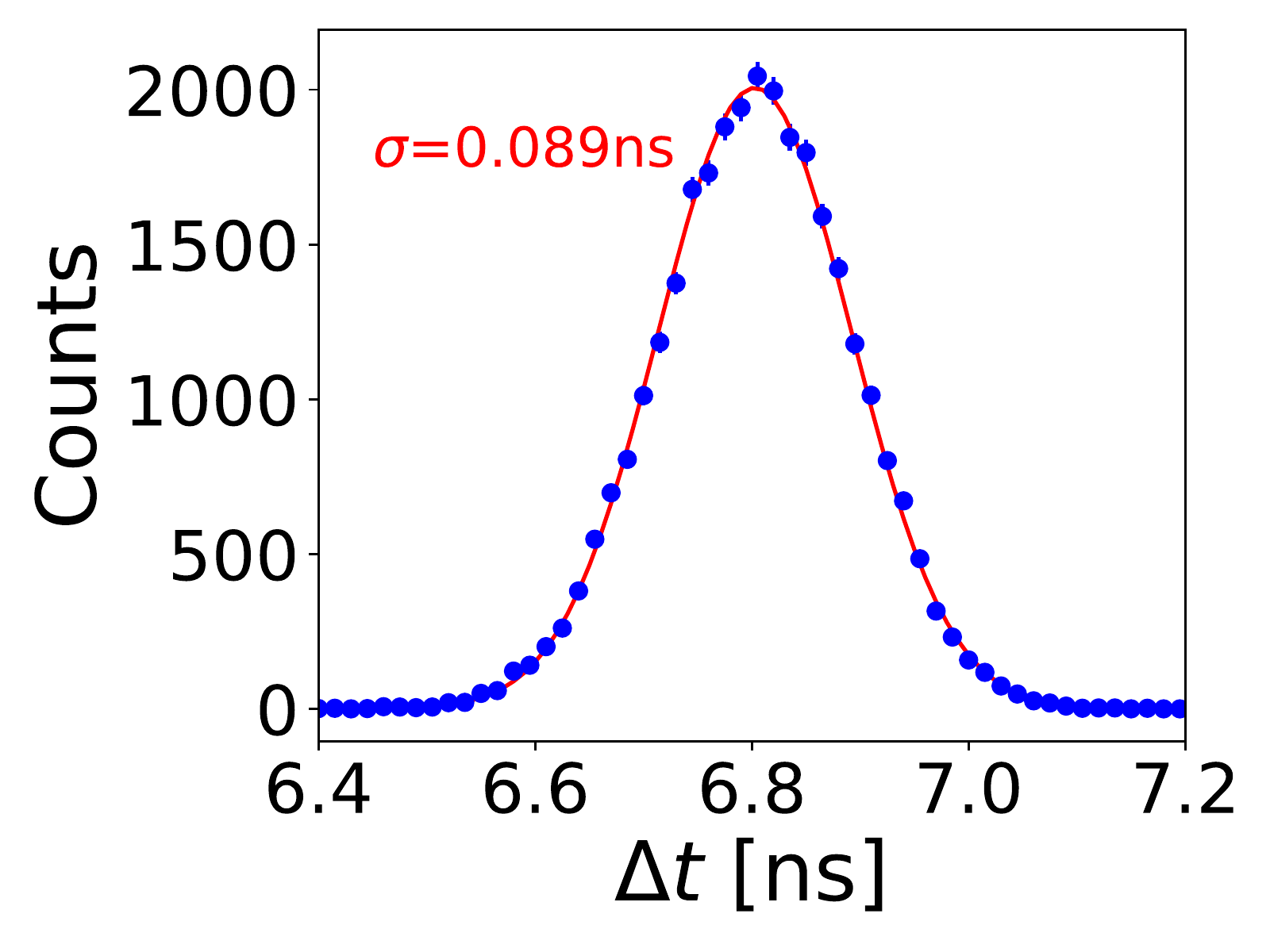}
\caption{$\Delta t$ spectrum measured with laser system.}
\label{fig:tres}
\end{figure}

A typical $\Delta t$ spectrum measured with the laser calibration system is shown in Figure~\ref{fig:tres}.  The mean and variance of a Gaussian fit to the spectrum yield the mean value of $\Delta t$ and time resolution, respectively.  Note that the time resolution is on the order of 100 ps.   

\subsection{Time walk correction}

One of the primary uses of the laser system is to allow a time walk correction (TWC) for the TDC.  Leading-edge discriminators apply a fixed threshold value to an input signal, above which the input signal is transformed into a logical NIM pulse output.  The timing of the output pulse corresponds to the time the input pulse passes the threshold, and thus become later for smaller pulses.  When a discriminator is used with a TDC, this leads to timing measurements that depend on the height of the pulse.  This is referred to as ``time walk", and must be corrected to achieve better time resolution, independent of the signal amplitude.  

\begin{figure*}[h!]
\centering  
\begin{subfigure}{\columnwidth}
\centering  
\includegraphics[width=\textwidth]{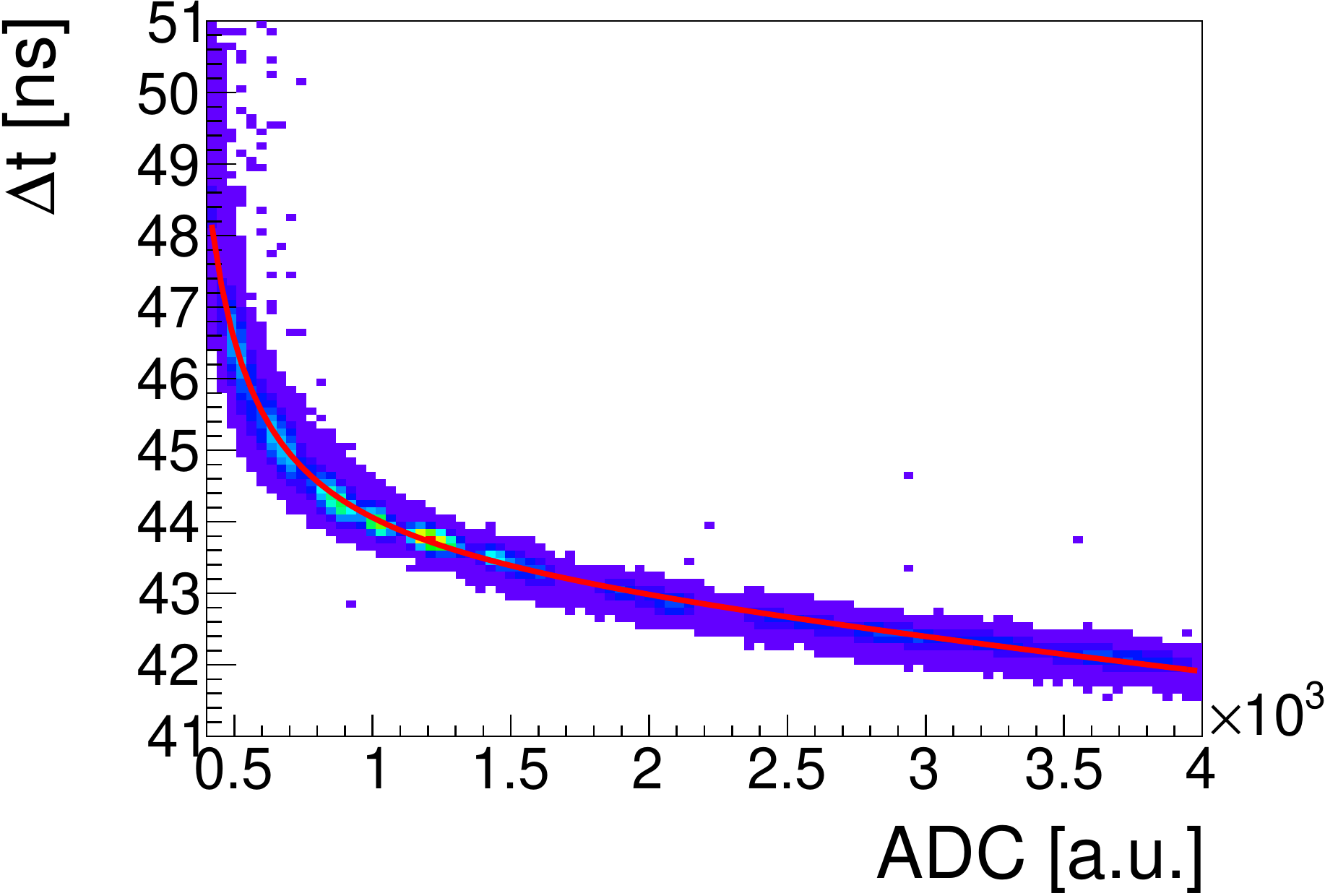}
\caption{}
\label{fig:fbBefore}
\end{subfigure}
\begin{subfigure}{\columnwidth}
\centering  
\includegraphics[width=\textwidth]{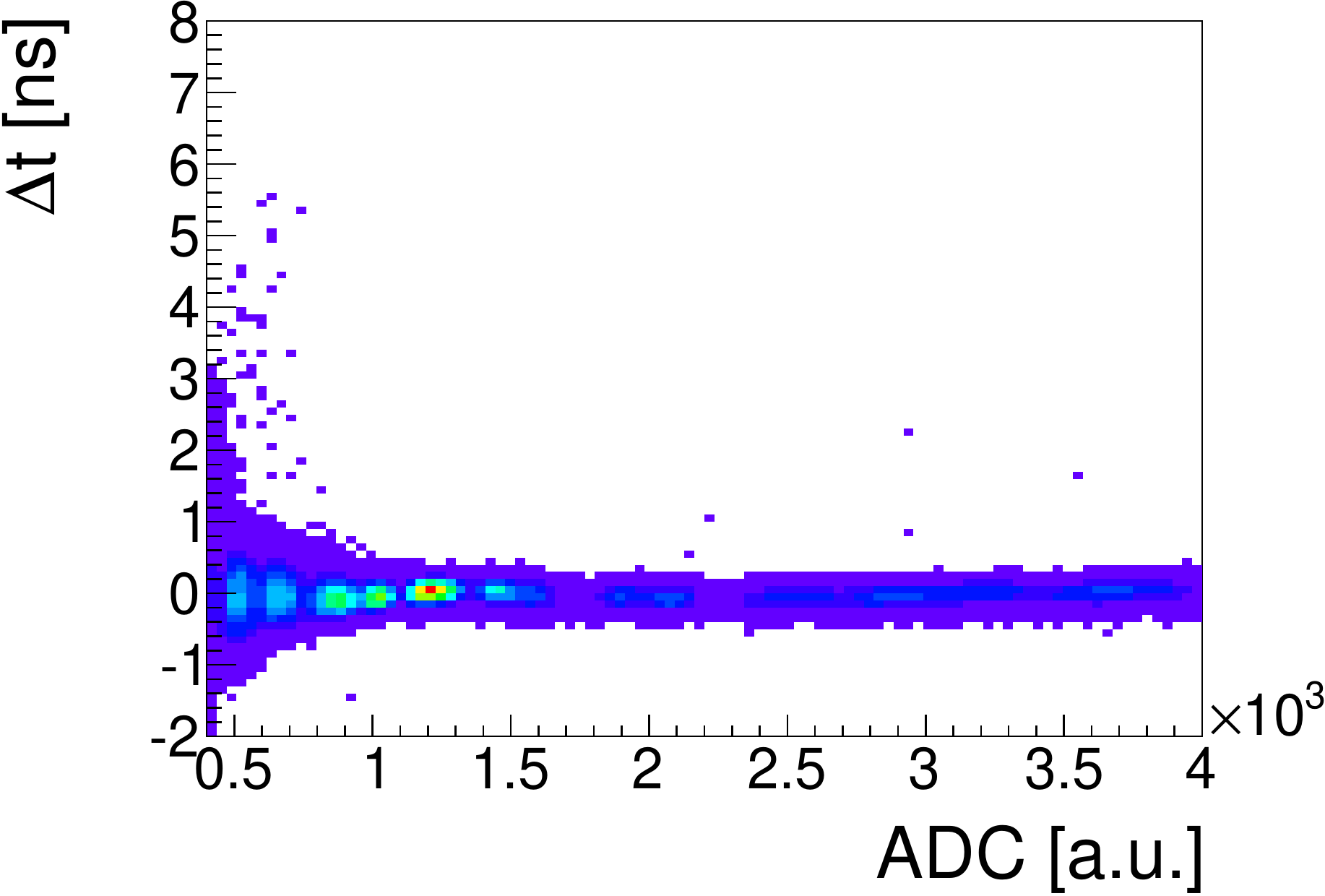}
\caption{}
\label{fig:fbAfter}
\end{subfigure}
\caption{(a) Initial time walk curve with with fit, showing the ADC dependence of the $\Delta t$ measurement. (b) Corrected time walk curve, showing no ADC dependence.}
\end{figure*}

The TWC is obtained by varying the attenuation of the laser pulses delivered to the scintillator, effectively varying the amplitudes of the discriminated waveforms.  This data can be used to produce a time walk curve (see Figure \ref{fig:fbBefore}) showing how timing measurement from the TDC changes as a function of the ADC signal $A$.  The curve is fit with the functional form
\begin{equation}\label{TWC}
\Delta t = \frac{1}{c_0 A + c_1} + c_2 A + c_3
\end{equation}
where the $\Delta t$ is defined by Eq.~\ref{eq:tof}.  The $c_i$ are free parameters of the fit.  For each event, this fit is used to correct the $\Delta t$ measurement as a function of the corresponding ADC signal.  The time walk corrected curve is shown in Figure \ref{fig:fbAfter}.  The strong dependence of $\Delta t$ on ADC signal is eliminated with the correction.  This calibration is performed individually for each scintillator in the detector.  By running the laser at a sufficiently high repetition rate ($\sim$kHz), this calibration can be performed in a matter of seconds. 

In addition to the TWC, this quick calibration also establishes an absolute reference time for all TDC channels. Because the fiber optic cables from the splitter are all of equal length, each scintillator receives its laser pulse at the same time. Therefore, all cable delays between the PMTs and electronics can be quickly accounted for by taking data with the laser system. 

\section{Drift corrections}

Over the course of an experiment, the performance of detectors and electronics can change or drift due to effects such as temperature fluctuations, radiation damage, or normal wear and tear.  Because the laser system maintains fixed timing throughout the experiment, it can be used to correct the effect these drifts have on timing measurements. 

To demonstrate this correction, the laser system was run at 500 Hz over the course of 1 hour. To simulate changes in timing due to drifts in the detectors or electronics, a delay was attached to the output signal of the PMT.  Initially, the delay was set to zero.  At times (10, 30, 40) minutes, the delay was set to (0.5, 1.0, 1.5) ns. 

\begin{figure*}[h!]
\centering
\begin{minipage}{0.99\columnwidth}
\centering
\includegraphics[width=\columnwidth]{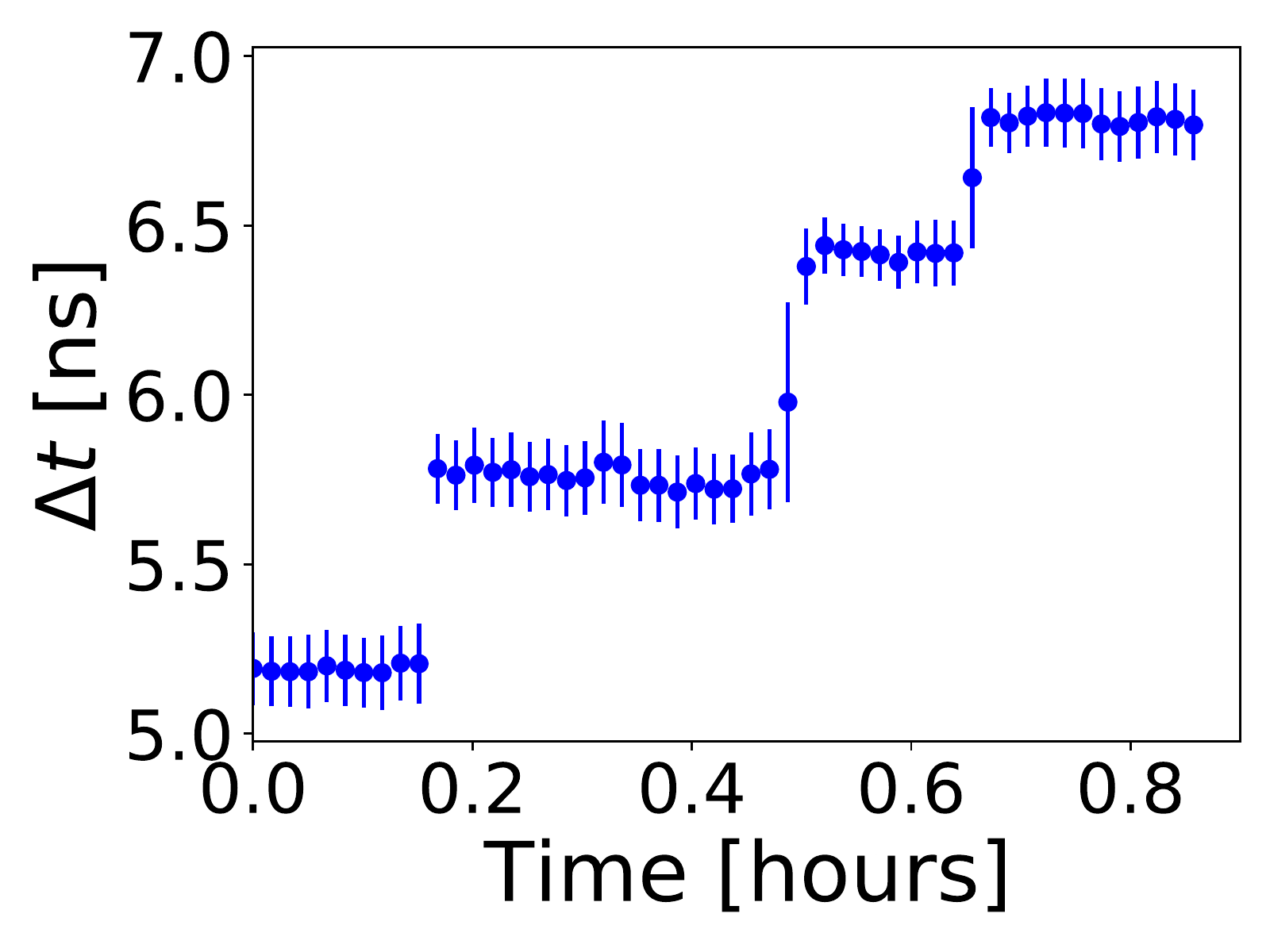}
\caption{$\Delta t$ as a function of time. Each blue data point is obtained from the $\Delta t$ spectrum over one minute of measurements. The central values and uncertainties of each point are obtained from a Gaussian fit to $\Delta t$, as shown in Figure~\ref{fig:tres}.}
\label{fig:tdcTracking}
\end{minipage}
\hspace{20pt}
\begin{minipage}{0.99\columnwidth}
\centering
\includegraphics[width=\columnwidth]{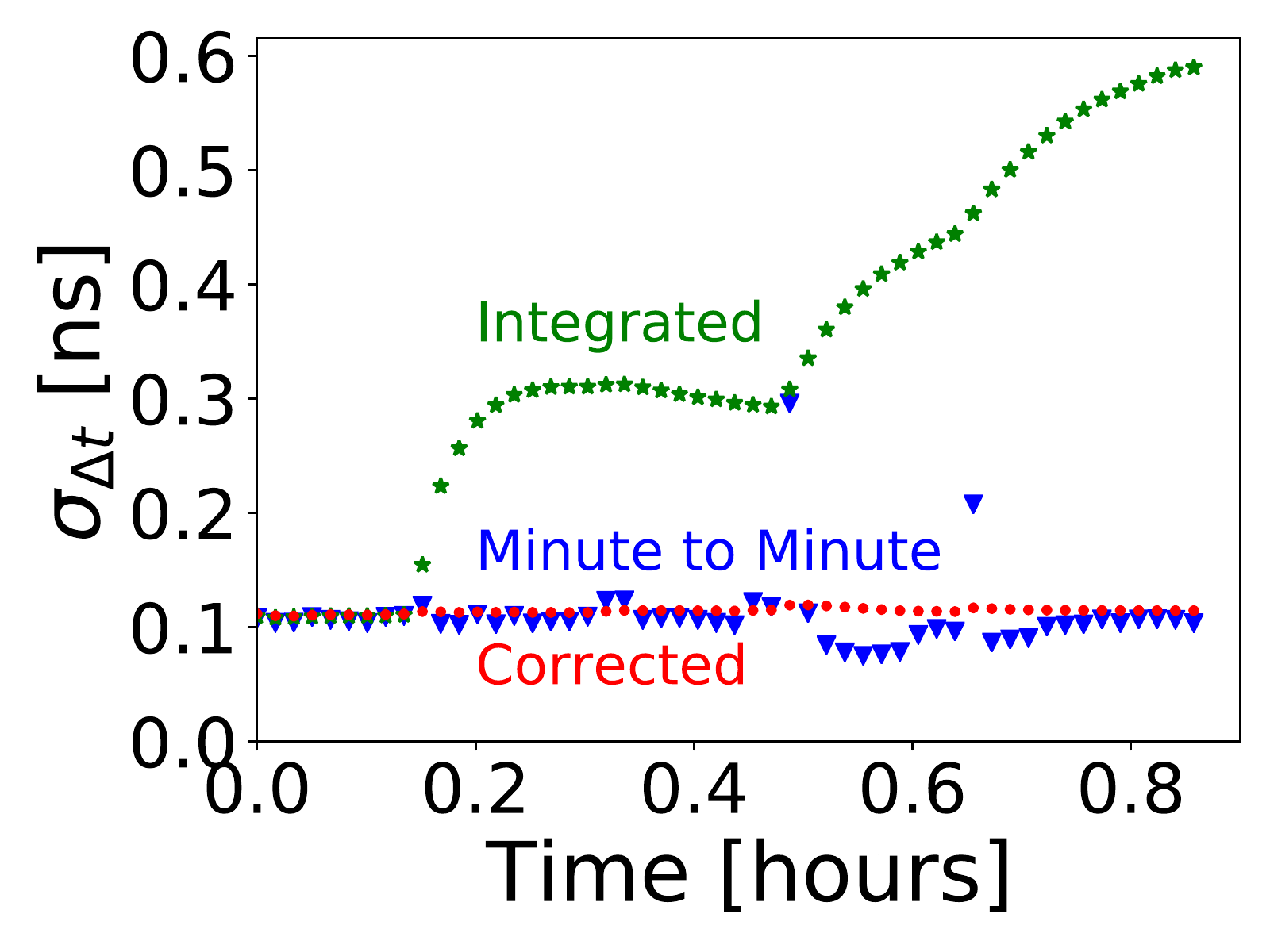}
\caption{$\Delta t$ resolution as a function of time.  The blue triangles show the resolution of each individual $\Delta t$ measurement. The green stars show the integrated $\Delta t$ resolution. The red circles show the integrated $\Delta t$ resolution, corrected for the drifts in $\Delta t$. }
\label{fig:tdcRes}
\end{minipage}
\end{figure*}

Figure \ref{fig:tdcTracking} shows how the delay changed the $\Delta t$ measurements over the course of the hour.  The resolution of each $\Delta t$ measurement is approximately constant in time. However, if the measurements are integrated over time, the total resolution becomes larger as the delays are increased.  The laser system can be used to correct for this change by subtracting the average $\Delta t$ minute by minute. Figure \ref{fig:tdcRes} shows that before the first delay, the total integrated resolution is nearly the same as the minute to minute resolution, approximately 0.1 ns.  As delays are added, the total integrated resolution increases. After the correction is made, the total integrated resolution remains the same as the minute to minute resolution.

\section{Laser signal stability}

It was observed that the amplitude of the laser pulse delivered to each scintillator changes over time.  The cause of this drift was isolated to the 1$\times$400 splitter. When the splitter is removed from the system (and the attenuator is used to replicate the amplitude reduction of the splitter), the amplitude remains constant.  Moreover, the effect was found to vanish when an identical test was performed with the 405 nm laser. 

\begin{figure}[b!]
    \centering
    \includegraphics[width=\columnwidth]{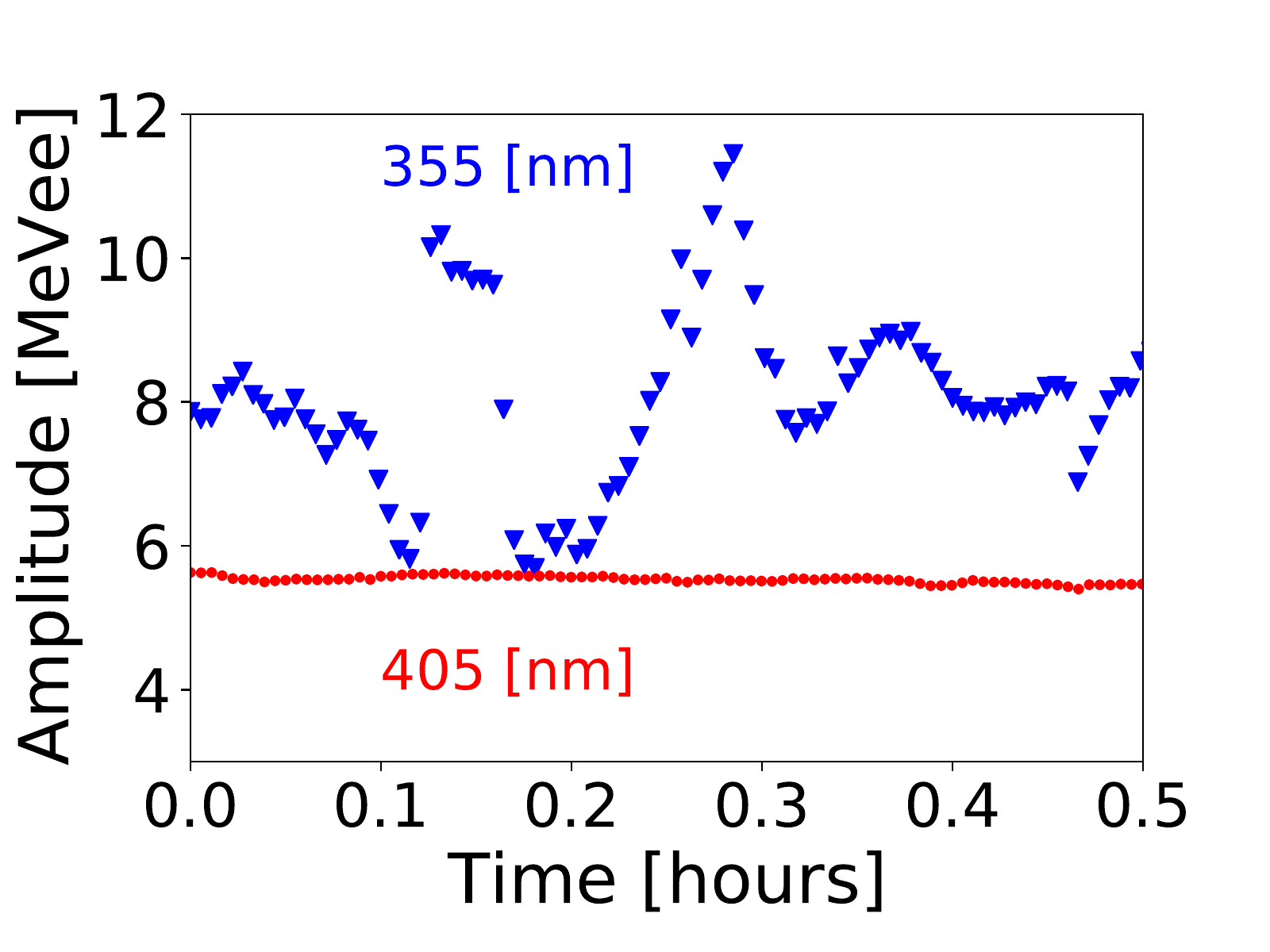}
    \caption{ADC signal as a function of time for the attenuated 355 nm laser (blue triangles) and 405 nm laser (red circles).}
    \label{fig:adcWand}
\end{figure}

As previously mentioned, the pulses from the 405 nm laser carried significantly less energy than the 355 nm laser pulses.  To correct for this energy difference, pulses from the 405 nm laser were compared to attenuated pulses from to the 355 nm laser. Despite the differences in pulse length and frequency, the PMT response from the induced scintillation is largely the same.  The ADC was monitored over the course of 1.25 hours at 500 Hz to track the changing amplitude of the system. Figure \ref{fig:adcWand} shows how the ADC changed for one scintillator over the course of the run. The setup with the 355 nm laser shows large wandering of the ADC amplitude.  In the setup with the 405 nm laser, the ADC amplitude remains constant.  The conversion from ADC channel to MeVee was calibrated using a $^{60}$Co source.

\section{Conclusion}

A laser calibration for TOF scintillator arrays was developed.  The system was tested with both a 1$\times$6 and 1$\times$400 fiber-optic splitter, allowing application to both small- and large-scale scintillator arrays.  The laser system was successfully implemented in BAND to provide a time walk correction, absolute time calibration, and TOF drift correction.  The ability to continuously monitor the laser pulse heights during data taking enables off-line corrections of both energy measurements and spacial distributions of the detected particles. The system currently has amplitude instabilities on the level of  ~$\pm$2\% (section 3.3)  to ~$\pm$30\% (section 7).  While this did not have any significant negative impact for BAND, which measures neutron momentum purely from time of flight, this must be improved for the system to be deployed in detectors where amplitude calibrations are also critical. One potential application is the LAD Experiment planned for JLab's Hall C~\cite{lad}, which will use the Large Acceptance Detector (LAD) to detect recoiling spectator protons. In LAD, amplitude information from the scintillators will be crucial for reducing accidental coincidence backgrounds. Therefore, it is a goal for future implementations to improve the amplitude stability of the system.

This research was supported by the U.S. Department of Energy, Office of Science, Office of Nuclear Physics under Award Numbers DE-FG02-94ER40818 and DE-SC0020240, the Israel Science Foundation under Grant Numbers 136/12 and 1334/16, the Pazy Foundation, and the Israel Atomic Energy Commission.

\bibliography{references}

\begin{thebibliography}{10}
\expandafter\ifx\csname url\endcsname\relax
  \def\url#1{\texttt{#1}}\fi
\expandafter\ifx\csname urlprefix\endcsname\relax\def\urlprefix{URL }\fi
\expandafter\ifx\csname href\endcsname\relax
  \def\href#1#2{#2} \def\path#1{#1}\fi

\bibitem{band}
E.~P. Segarra, et~al., \href{https://arxiv.org/abs/2004.10339}{The {CLAS12
  Backward Angle Neutron Detector (BAND)}}.
\newline\urlprefix\url{https://arxiv.org/abs/2004.10339}

\bibitem{Hen:2016kwk}
O.~Hen, G.~Miller, E.~Piasetzky, L.~Weinstein, {Nucleon-Nucleon Correlations,
  Short-lived Excitations, and the Quarks Within}, Rev. Mod. Phys. 89~(4)
  (2017) 045002.
\newblock \href {http://arxiv.org/abs/1611.09748} {\path{arXiv:1611.09748}},
  \href {http://dx.doi.org/10.1103/RevModPhys.89.045002}
  {\path{doi:10.1103/RevModPhys.89.045002}}.

\bibitem{teem_laser}
\text{Teem Photonics STV-01E-140 pulse laser}, \newline
  \url{https://www.teemphotonics.com/laser/microchip-laser-355-nm-1e-4khz-600ps}.

\bibitem{Smith:1999ii}
E.~Smith, et~al., {The time-of-flight system for CLAS}, Nucl.\ Instrum.\ Meth.\
  A 432 (1999) 265--298.
\newblock \href {http://dx.doi.org/10.1016/S0168-9002(99)00484-2}
  {\path{doi:10.1016/S0168-9002(99)00484-2}}.

\bibitem{Hasell:2009zza}
D.~Hasell, et~al., {The BLAST experiment}, Nucl.\ Instrum.\ Meth.\ A 603 (2009)
  247--262.
\newblock \href {http://dx.doi.org/10.1016/j.nima.2009.01.131}
  {\path{doi:10.1016/j.nima.2009.01.131}}.

\bibitem{thor_laser}
\text{Thorlabs NPL41B pulsed laser}, \newline
  \url{https://www.thorlabs.com/thorproduct.cfm?partnumber=NPL41B}.

\bibitem{thor_coupler}
\text{Thorlabs PAF-X-18-PC-A FiberPort coupler}, \newline
  \url{https://www.thorlabs.com/thorproduct.cfm?partnumber=PAF-X-18-PC-A}.

\bibitem{thor_mirror}
\text{Thorlabs NB1-K08 mirror}, \newline
  \url{https://www.thorlabs.com/thorproduct.cfm?partnumber=NB1-K08}.

\bibitem{newport_mode}
\text{Newport FM-1 fiber mode scrambler}, \newline
  \url{https://www.newport.com/p/FM-1}.

\bibitem{teem_control}
\text{Teem Photonics MLC-03A-DP1 laser controller}, \newline
  \url{https://www.teemphotonics.com/laser/mlc-03a-xp1/}.

\bibitem{siglent_pulse}
\text{Siglent SDG 1032X pulse generator}, \newline
  \url{https://siglentna.com/product/sdg1032x/}.

\bibitem{thor_photo}
\text{Thorlabs DET025AFC/M fiber-coupled silicon photodetector}, \newline
  \url{https://www.thorlabs.com/thorproduct.cfm?partnumber=DET025AFC/M}.

\bibitem{oz_attenuator}
\text{OZ Optics custom attenuator}, \newline
  \url{https://www.ozoptics.com/products/attenuators.html}.

\bibitem{lfiber_splitter}
\text{Lfiber fiber optic splitter}, \newline
  \url{https://www.lfiber.com/large-core-fiber-coupler-splitter/}.

\bibitem{thorlabs_splitter}
\text{Thorlabs fiber optic splitter}, \newline
  \url{https://www.thorlabs.com/navigation.cfm?guide_id=2211}.

\bibitem{hamamatsu_pmt}
\text{Hamamatsu R7724 PMT}, \newline
  \url{https://www.hamamatsu.com/jp/en/product/type/R7724/index.html}.

\bibitem{lecroy_fifo}
\text{Teledyne LeCroy 428F linear FIFO}, \newline
  \url{https://teledynelecroy.com/lrs/dsheets/428.htm}.

\bibitem{lecroy_disc}
\text{Teledyne LeCroy 623B octal discriminator}, \newline
  \url{https://teledynelecroy.com/lrs/dsheets/623.htm}.

\bibitem{phillips_gate}
\text{Phillips Scientific 794 gate generator}, \newline
  \url{http://www.phillipsscientific.com/preview/794pre.htm}.

\bibitem{caen_tdc}
\text{CAEN V1290A TDC}, \newline
  \url{https://www.caen.it/products/v1290a-2esst/}.

\bibitem{caen_adc}
\text{CAEN V792 ADC}, \newline \url{https://www.caen.it/products/v792/}.

\bibitem{lad}
O.~Hen, L.~Weinstein, S.~Wood, S.~Gilad, {In Medium Nucleon Structure
  Functions, SRC, and the EMC effect, Jefferson Lab experiment E12-11-107}
  (2011).

\end{thebibliography}

\end{document}